\documentclass[11pt]{article}
\usepackage{graphicx}

\usepackage[T1]{fontenc}
\usepackage{ae,aecompl}

\begin{document}
\title{\Large{{\bf Topological classification of RNA structures\\}}}
\vspace{1.5cm}
\renewcommand\thefootnote{\relax}
\author{~\\{\sc Michael Bon}$^{1,2,*}$, {\sc Graziano
Vernizzi}$^{3,*,\ddag}$, {\sc Henri
Orland}$^1$ and {\sc A. Zee}$^{4,5}$\\~\vspace{1cm}\\
$^1$Service de Physique Th\'eorique, CEA Saclay, 91191
Gif-sur-Yvette Cedex, France\\~\\
$^2$Ecole Nationale Sup\'erieure des Mines de Paris, 75006 Paris,
France\\~\\
$^3$Department of Materials Science and Engineering, Northwestern
University,\\ Evanston, IL 60208, USA\\~\\
$^4$Department of Physics, University of California, Santa Barbara,
CA 93106, USA\\~\\
$^5$Kavli Institute for Theoretical Physics, University of
California, Santa Barbara, CA 93106, USA\\~\\
} \maketitle  \footnotetext[1]{{\it Email addresses:}
\texttt{Michael.Bon@cea.fr} (Michael Bon),
\texttt{g-vernizzi@northwestern.edu} (Graziano Vernizzi),
\texttt{Henri.Orland@cea.fr} (Henri Orland),
\texttt{zee@kitp.ucsb.edu} (A. Zee).} \footnotetext[1]{$^*$These
authors contributed equally to this work.}
\footnotetext[1]{$^\ddag$To whom correspondence should be
addressed.}
\begin{abstract}
\textbf{We present a novel topological classification of RNA
secondary structures with pseudoknots. It is based on the
topological genus of the circular diagram  associated to the RNA
base-pair structure. The genus is a positive integer number, whose
value quantifies the topological complexity of the folded RNA
structure. In such a representation, planar diagrams correspond to
pure RNA secondary structures and have zero genus, whereas non
planar diagrams correspond to pseudoknotted structures and have
higher genus. We analyze real RNA structures from the databases wwPDB
and Pseudobase, and classify them according to their topological
genus. We compare the results of our statistical survey with
existing theoretical and numerical models. We also discuss possible
applications of this classification and show how it can be used for
identifying new RNA structural motifs. }
\end{abstract}


 \vfill
\newpage

\section*{Introduction}
In their biologically active form, RNA molecules are folded in
fairly well defined three dimensional structures \cite{PDB}. These
structures are strongly constrained by the pairing of conjugate
bases along the sequence, but depend also on the ionic strength of
the solution \cite{MD}. It has proved very useful to describe the
pairing of RNA in terms of secondary structures and pseudoknots
\cite{PRB}. These structural elements can be viewed as motifs which
appear repeatedly in the folds. The main structural motifs of
secondary structures are helical duplexes, single stranded regions,
hairpin stems, hairpin loops, bulges and internal loops, junctions
and multiloops (see table \ref{secondary}). It is convenient at this
stage to introduce some standard graphical representations of RNA
structures. In the {\it linear representation}, one writes the base
sequence on an oriented straight line, starting from the 5' to the
3' end. By replacing the straight line by a closed circle one
obtains the {\it circular representation}. The pairing of two bases
is represented by a dotted line, or colored line, joining the two
bases in the upper side of the straight 5'-to-3' line. In the case
of a circular representation, pairings are drawn inside the circle.
This representation associates a unique diagram to any set of base
pairings of RNA.
\begin{table}
\centering
\includegraphics[width=0.8\textwidth]{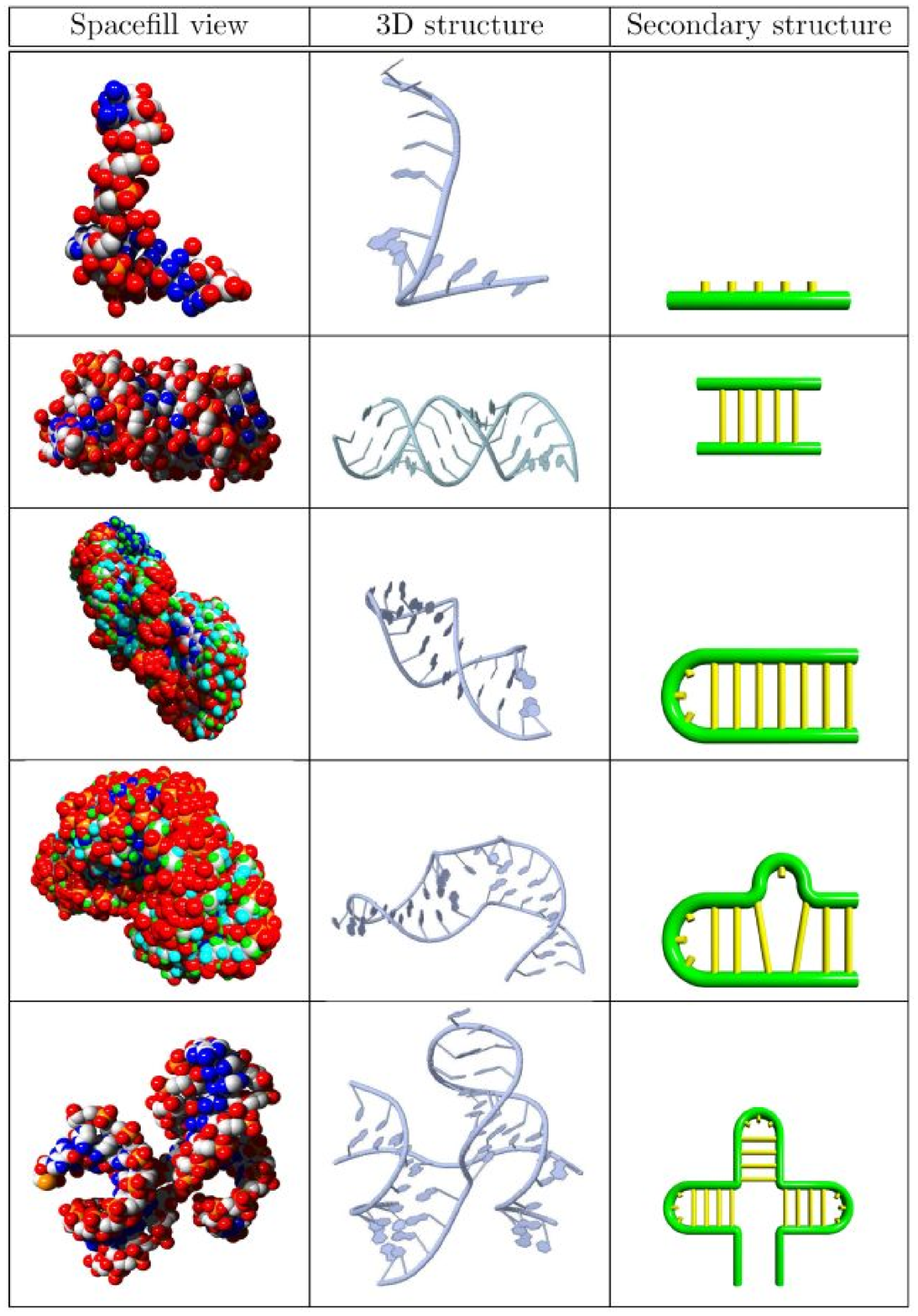}
\caption{Examples of basic RNA secondary structure motifs. From top
to bottom: a single strand (PDB 283D \cite{pdb_283d}), a helical
duplex (PDB 405d \cite{pdb_405d}), a hairpin stem and loop (PDB 1e4p
\cite{pdb_1e4p}), a bulge (PDB 1r7w \cite{pdb_1r7w}), a multiloop
(PDB 1kh6 \cite{pdb_1kh6}). From left to right: spacefill view,
three-dimensional structure, secondary structure motif. The pictures
are made with MolPov \cite{MolPov}, Jmol \cite{Jmol} and PovRay
\cite{PovRay}.} \label{secondary}
\end{table}
In the circular and linear representation, a diagram represents a
secondary structure if it involves only pairings which do not cross
\cite{W_sec}. In table \ref{secstruc} (top row), we show a secondary
structure, together with its two representations (linear and
circular in the fourth and fifth column, respectively). Similarly, a
diagram contains a pseudoknot if it contains pairings which do cross
( see, e.g., the bottom row in table \ref{secstruc}).
\begin{table}
\centering
\includegraphics[width=0.9\textwidth]{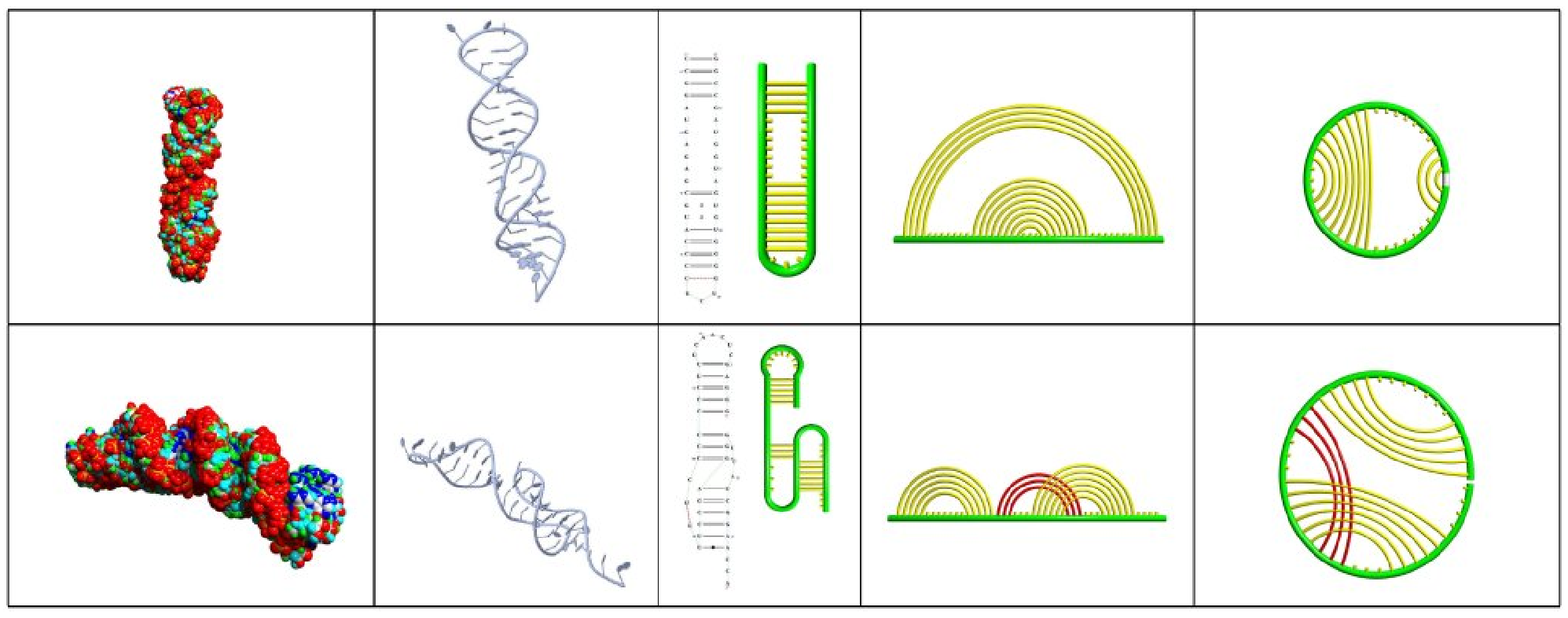}
\caption{Top row: example of a secondary structure motif  (a helix,
PDB 1a51 \cite{pdb_1a51}). Bottom row: an example of a common RNA
H-pseudoknot (PDB 1a60 \cite{pdb_1a60}). From left to right:
spacefill view, three-dimensional structure, secondary structure
(and base-pairings from RNAView \cite{RNAview}), linear
representation and circular representation. In red are emphasized
the non-planar pairings (crossing arcs).} \label{secstruc}
\end{table}

There are quite a few methods to predict secondary structures.
Energy-based methods have proven to be the most reliable (as, e.g.,
\cite{zuker,ViennaPackage}). They assign some energy to the base
pairings and some entropy to the loops and bulges. In addition, they
take into account stacking energies, and assign precise weights to
specific patterns (tetraloops, multiloops, etc.) \cite{david}. The
lowest free energy folds are obtained either by dynamic programming
algorithms \cite{zuker2}, or by computing the partition function of
the RNA molecule \cite{partition}. The main drawback of these
energy-based methods  is that they deal solely with secondary
structures and cannot take into account pseudoknots in a systematic
way.

There are several computer programs that attempt to predict
RNA-folding with pseudoknots, but the problem is still mostly
unsolved (see, e.g. \cite{PseudoPrograms,P1,P2,P3,P4,P5,P6,P7,P8};
the list is not exhaustive)  . There exists however a novel
approach: in order to include the pseudoknots, the RNA folding
problem has been formulated in terms of a sophisticated mathematical
theory, namely a quantum matrix field theory \cite{OZ}. These types
of field theories were first introduced in particle physics, more
precisely in Quantum Chromodynamics, in order to model the theory of
strong interactions \cite{thooft}. Since then, these field theories
have been used in many mathematical problems, such as combinatorics,
number theory, etc. (for a recent review see \cite{MatrixModels}).
They involve a parameter $N$, the linear dimension of the $N\times
N$ matrices, which can be used as an expansion parameter for the
theory (large $N$ expansion) \cite{thooft}. In the RNA folding
problem, the matrix field theory can be expanded diagrammatically in
various parameters. The simplest development is in terms of the
number of pairings and can easily be represented in terms of
diagrams. These diagrams, which are the usual Feynman diagrams of
quantum field theory, can be viewed as the set of all the possible
pairings of the RNA, with the correct corresponding Boltzmann
weights \cite{OZ,Z}. Another possible expansion is in powers of
$1/N$. As was shown in a previous paper \cite{OZ,VO}, this expansion
relies on a topological number called the genus which characterizes
the pairing. As we shall see, the genus of a diagram is defined by
its  embedding on a two-dimensional surface. It is the minimal
number of handles that the surface should have so that the diagram
can be drawn on the surface without crossing.

Secondary structures correspond to zero genus, that is planar
structures: They can be drawn on a sphere without crossing. The
simplest pseudoknots, such as the ``H-pseudoknot'' (see table
\ref{secstruc}) or the kissing hairpin, correspond to genus 1: they
can be drawn on a torus without crossing. This classification of RNA
structures allows us to completely grasp the topological complexity
of a pseudoknot with a single integer number, the genus. It can be
viewed as a kind of ``quantum'' number. It is reminiscent of the
superfold families, such as CATH or SCOP \cite{protein}, which have
proven so useful in protein structure classification. In the
literature other possible classifications of RNA structures with
pseudoknots have been proposed, such as the ones in, e.g. ,
\cite{others}. However, the one we propose in this paper is the only
one that is purely topological, i.e. independent of any
three-dimensional embedding and which is based only on the classical
topological expansion of closed bounded surfaces. This is also the
reason why this expansion can be derived mathematically with
standard tools of combinatorial topology. We believe that such a
mathematical framework can be exploited far beyond the simple
classification of RNA pseudoknots, and could be applied also for
RNA-folding predictions \cite{VO}. In this work however we restrict
only to the problem of classifying known RNA-structures.

In the following, we shall define more precisely the genus for a
given diagram, and show how it can be simply calculated. We then
present an analysis of the genii of two main databases which contain
RNA structures, namely PSEUDOBASE \cite{Pseudobase} and the wwPDB
(the Worldwide Protein Data Bank which contains some RNAs). The RNA
structures in the latter are also listed in the RNABase database
\cite{RNAbase}, that we also used as a reference database. We find
that RNAs of sizes up to about 250 have a genus smaller than 2,
whereas long RNAs, such as ribozomal RNA may have a genus up to 18.

\section*{Materials and Methods}

\begin{figure}[hbpt]
\centering
\includegraphics[width=0.48\textwidth]{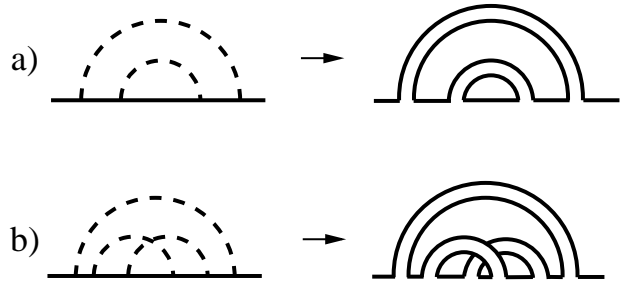}
\caption{A schematic view of the double line representation (right)
of a generic linear representation of pairings (left). The example a)
(top) represents a couple of stacked base-pairs, and b) (bottom)
represents an H-pseudoknot embedded in an hairpin.} \label{double}
\end{figure}

\underline{The genus}. The topological classification of RNA
secondary structures with pseudoknots that we propose is based on
the concept of topological {\it genus}. We first review the
definition of genus of a given diagram. Consider a diagram
representing a pairing in the linear representation. The matrix
field theory representation of the problem suggests representing a
pairing not by a single dotted line, but rather by a double line
(which should never be twisted) \cite{OZ,thooft}. Therefore, a
unique diagram in the double line representation corresponds to each
dotted-line diagram. Some examples are shown in fig.\ref{double}.

\begin{figure}[hbpt]
\centering
\includegraphics[width=0.48\textwidth]{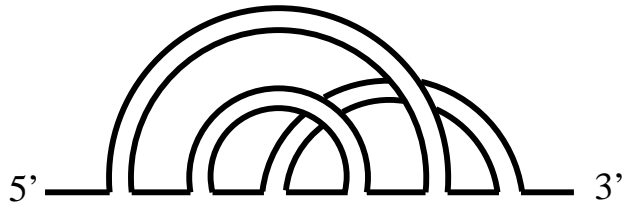}
\caption{This diagram represents a pseudoknot with genus $g=1$ since
it has $P=3$ double lines and $L=1$ loops.} \label{ex1}
\end{figure}

Each double line diagram is characterized by its number of double
lines (i.e. the number of pairings of the diagram) which we denote
by $P$, and by its number of loops denoted $L$, which is the total
number of closed loops made with the (single) lines of the diagram.
For instance, in fig.\ref{double} (bottom) and in fig.\ref{ex1}, the
diagram has $P=3$ double lines and $L=1$ loop. The genus of the
diagram is the integer defined by
\[ g=\frac{P-L}{2} \]

\begin{figure}[hbpt]
\centering \framebox{
\includegraphics[width=0.60\textwidth]{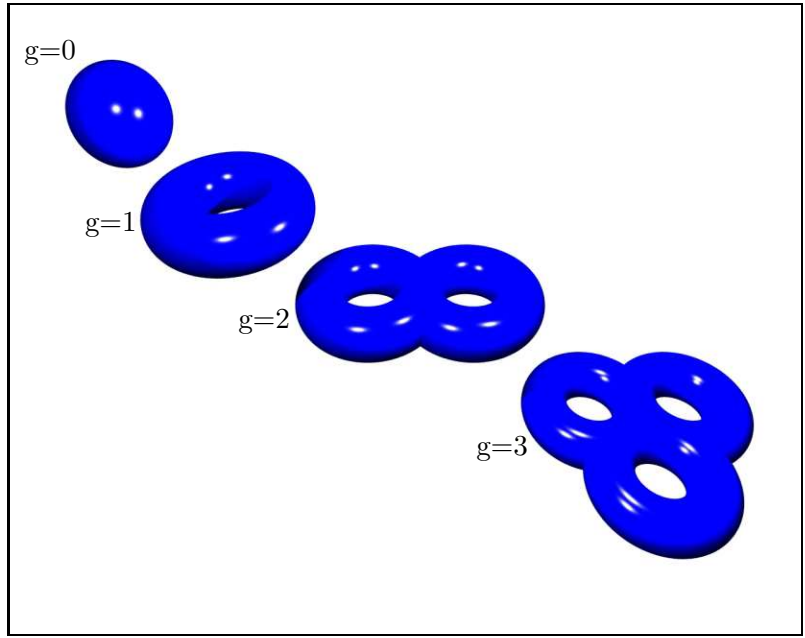}
\put(-291,215){g=0} \put(-268,150){g=1} \put(-210,113){g=2}
\put(-120,65){g=3}} \caption{First few terms of the topological
expansion of closed oriented surfaces: the term $g=0$ is a sphere,
$g=1$ is a torus, $g=2$ is a double torus and so forth.} \label{gex}
\end{figure}

\begin{figure}[hbpt]
\centering \framebox{
\includegraphics[width=0.80\textwidth]{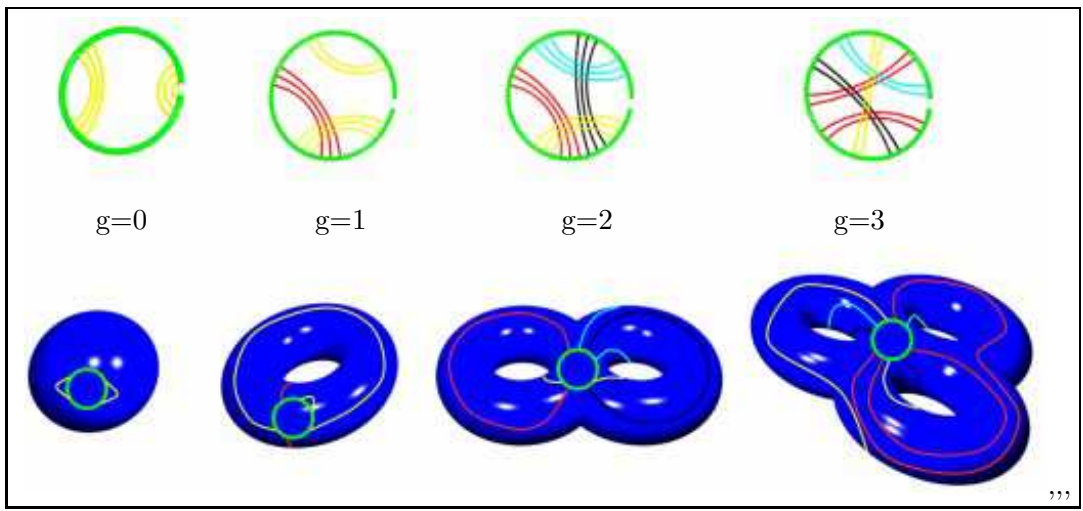}
\put(-360,100){g=0},\put(-280,100){g=1},\put(-190,100){g=2},\put(-90,100){g=3}}
\caption{Any RNA circular diagram can be drawn on a closed surface
with a suitable number of ``handles'' (the genus). For the sake of
simplicity, in this figure all helices and set of pairings on the
surfaces are schematically identified only by their color. Note that
the circle of the RNA-backbone (in green) topologically corresponds
to a hole (or puncture) on the surface.} \label{ge0}
\end{figure}

\begin{figure}[hbpt]
\centering \framebox{
\includegraphics[width=0.50\textwidth]{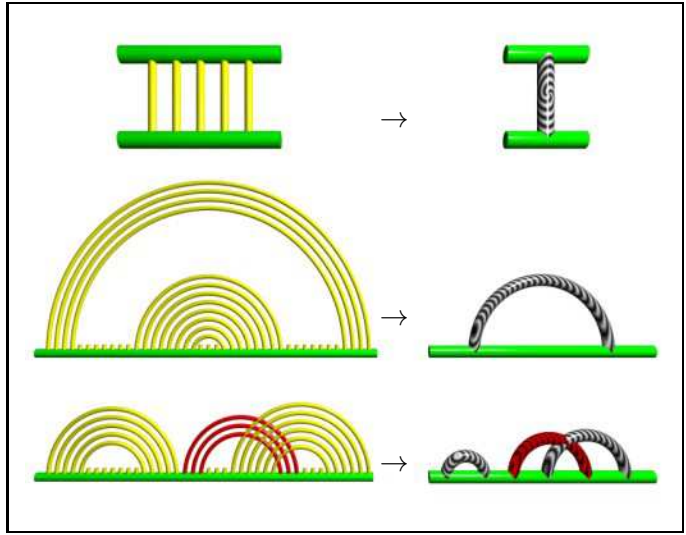}
} \put(-115,20){$\rightarrow$}
\put(-115,75){$\rightarrow$}\put(-115,150){$\rightarrow$}
\caption{The genus of a diagram does not change by identifying a
stack of paired bases with a single {\it effective} base-pair.}
\label{parallel}
\end{figure}

\begin{figure}[hbpt]
\centering\framebox{
\includegraphics[width=0.70\textwidth]{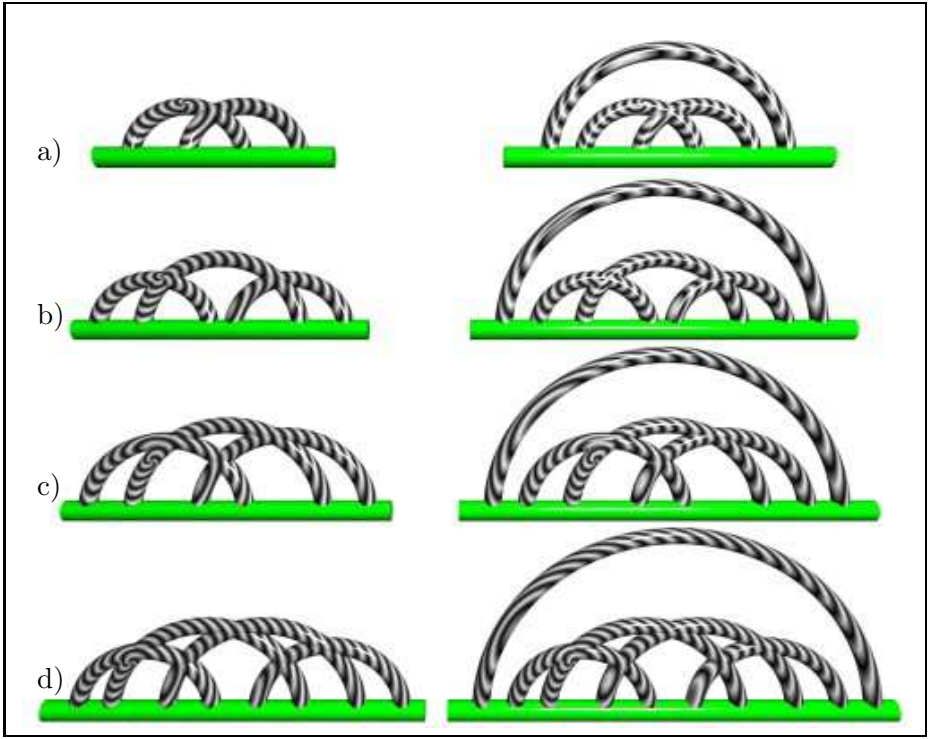}
\put(-333,215){a)} \put(-333,153){b)} \put(-333,88){c)}
\put(-333,15){d)}
 } \caption{These are the only 8 types of irreducible
pseudoknots with genus $g=1$.} \label{genus1}
\end{figure}

It is related to the Euler characteristics of the diagram, and is a
topological invariant of the diagram. Its geometrical interpretation
is quite simple. Consider a sphere with $g$ handles: a sphere with 0
handles is a sphere, a sphere with one handle is topologically
equivalent to a torus, a sphere with 2 handles is topologically
equivalent to a double-torus, etc. (see fig.\ref{gex}).
The genus $g$ of a diagram is the minimum number of handles a sphere
must have in order to be able to draw the diagram on it without any
crossing. The precise way to do so, is unambiguously defined only
when the diagram does not have open dangling lines, such as the $5'$
or $3'$ ends. Therefore it is important to connect the ends, as is
done in the circular representation. However, it is more convenient
to close the two ends {\it below} the backbone-line, which results
in drawing the pairing arcs all at the exterior of the
backbone-circle. In that way it is simple to see how the embedding
of a pseudoknotted RNA structure on a high-genus surface works.
Mathematically speaking, the circle of the RNA-backbone (when the
$5'$ are $3'$ connected) becomes the boundary of a hole or {\it
puncture} on the surface, and the arcs corresponding to the RNA
base-pairs are drawn on the surface without that hole. In
fig.\ref{ge0}, we show explicit examples of diagrams having
different genus. As can be seen, a diagram with genus 0 is planar,
in that it can be drawn on the sphere without crossing, and
corresponds to a secondary structure. More generally, it was shown
in \cite{OZ} that the secondary structure diagrams are all the
planar diagrams with $g=0$. Likewise, in fig.\ref{ge0} one sees also
how diagrams with non-zero genus $g\neq0$ can be drawn without any
crossing on a surface with $g$ handles.
Clearly, different diagrams can have the same genus. Thus, in order
to further simplify the classification, we first note that adding a
line of pairing parallel to an existing one does not change the
genus of the diagram, since it increases by one the number of
pairing lines, and increases by one the number of loops of the
diagram. Therefore, all diagrams with parallel pairings are
equivalent topologically. We will thus use a reduced representation
of the diagrams, where each pairing line can be replaced by any
number of parallel pairings as in fig.\ref{parallel}.
With this convention, it has been shown in \cite{poz} that there are
exactly 8 topologies of pseudoknots of genus 1, see
fig.\ref{genus1}. Those topologies can be uniquely identified also
as a) ABAB, b) ABACBC, c) ABCABC, d) ABCADBCD, where each letter
A,B, etc. indicates a specific helix (or set of helices) along the
RNA-backbone from the $5'$ end to the $3'$ end. Note that one
recognizes the standard H-pseudoknot (ABAB) and the kissing hairpin
(ABACBC) (diagrams a) and b) on the left of fig.\ref{genus1},
respectively). Among the 8 pseudoknots of genus 1, four are quite
common in the databases (the rows a) and b) of fig.\ref{genus1}),
two are very rare (the row c) fig.\ref{genus1}), and the remaining
two have not been reported as of yet. We will discuss these
pseudoknots in more details in the next section.
Let us insist again that the genus captures the topological
complexity of the pseudoknots. It is not simply related to the
number of crossings, or of pairings. It depends on the intrinsic
complexity of the pseudoknot. This complexity itself depends on what
kind of pairings are considered. This is of course conventional.
Before discussing the statistics of the genus of pseudoknots from
the databases, let us address this question. As discussed in
\cite{West}, there are many possible non-canonical bonds between
base-pairs. We emphasize that our classification of RNA structures
according to  their genus is well defined and possible even when
including non-canonical bonds, or more general definitions of
RNA-binding interactions (as far as such interactions are binary).
The larger the number of pairings, the higher the genus of the
structure might be. However, the weaker bonds, such as the Hoogsteen
bonds, or even the wobble pairs, do not form the structure, they
merely stabilize a structure already formed by canonical pairings.
Therefore, in the following, we shall consider only Watson-Crick
pairs between conjugate bases and G-U wobble pairs.

\begin{figure}[hbpt]
\centering \framebox{
\includegraphics[width=0.48\textwidth]{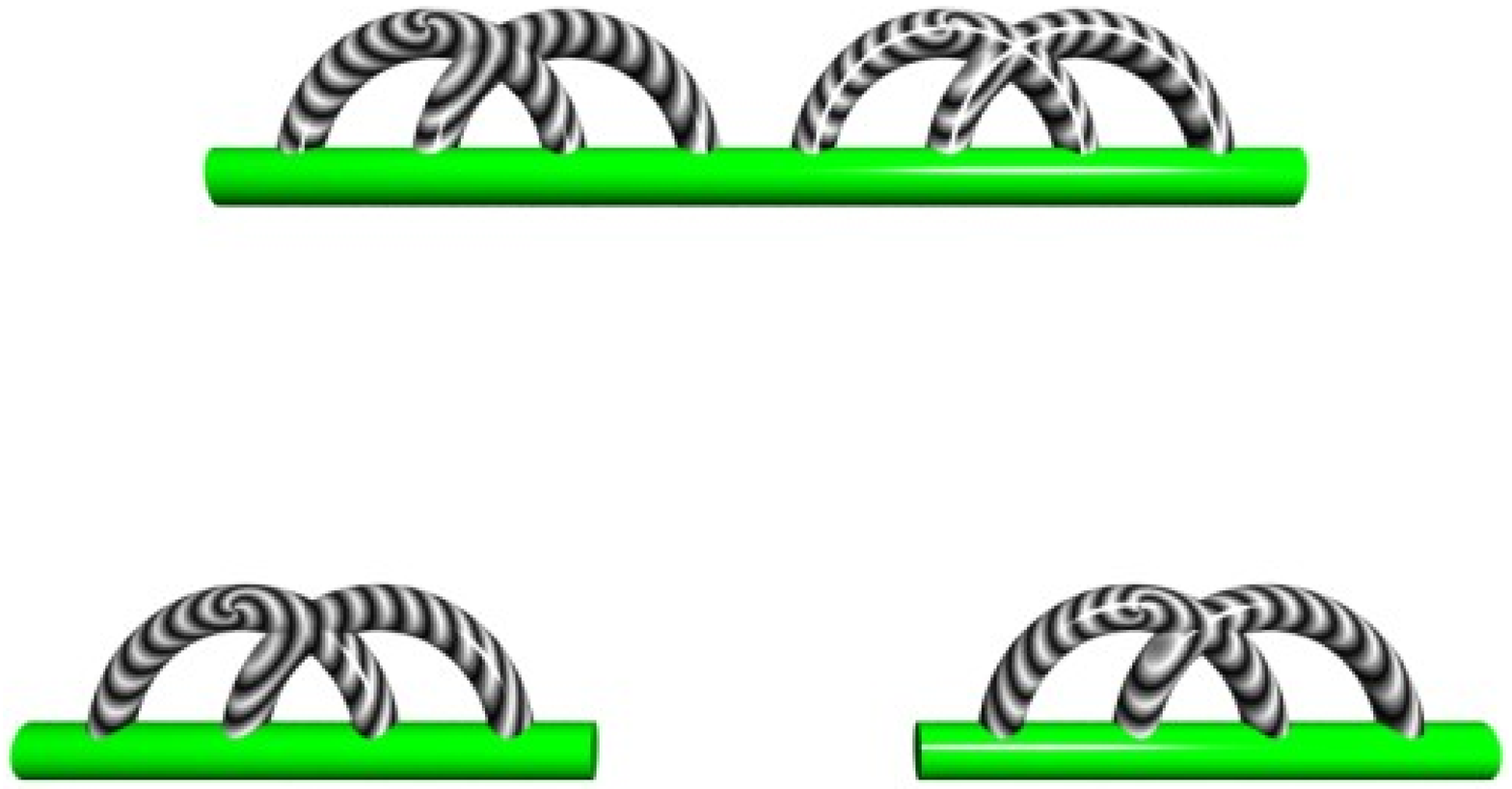}
\put(-130,100){\vector(-1,-1){10}} \put(-100,100){\vector(1,-1){10}}
} \framebox{\includegraphics[width=0.38\textwidth]{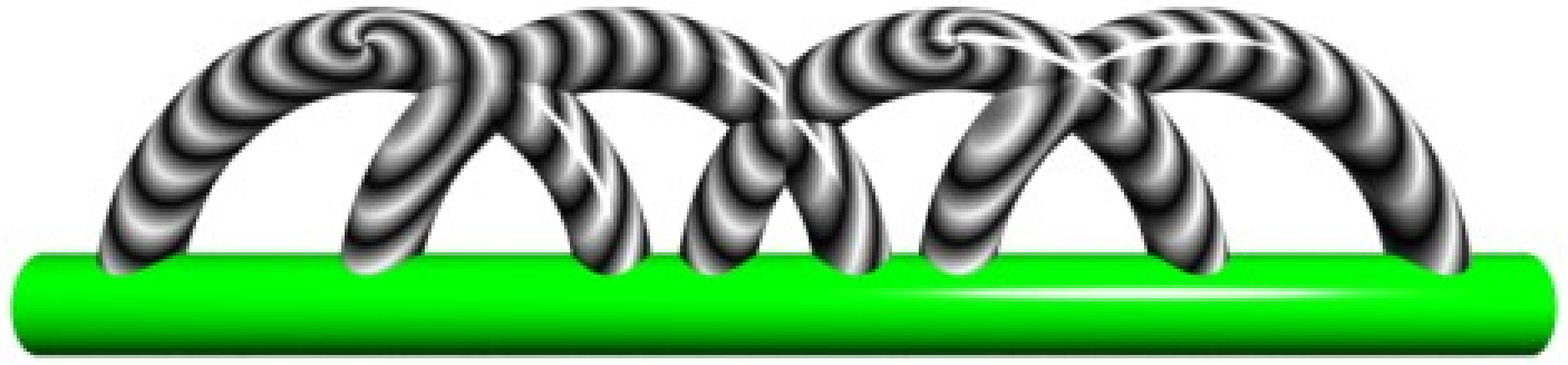} }
\caption{Example of a reducible pseudoknot (left) and an irreducible
one (right). The reducible pseudoknot can be split in two
disconnected parts, as shown, by cutting the backbone only once. The
total genus is the sum of the genus of the two components (in this
example the total genus is 2).} \label{irrnonirr}
\end{figure}

\underline{Irreducibility and nesting}. In many cases, the genus of
a diagram is an additive quantity. For instance, if we consider a
succession of two H-pseudoknots (see fig.\ref{irrnonirr}, left),
each one has genus 1, and the total genus of the diagram is 2. In
order to characterize the intrinsic complexity of a pseudoknot, it
is thus desirable to define the notion of {\it irreducibility}. A
diagram is said to be irreducible if it can not be broken into two
disconnected pieces by cutting a single line. The diagram on the
left of fig.\ref{irrnonirr} is reducible, whereas the one on the
right of fig.\ref{irrnonirr} is irreducible. Any diagram can thus be
decomposed in a unique way into irreducible parts. It is obvious
that the genus of a non-irreducible diagram is the sum of the genii
of its irreducible components.

\begin{figure}[hbpt]
\centering \framebox{
\includegraphics[width=0.40\textwidth]{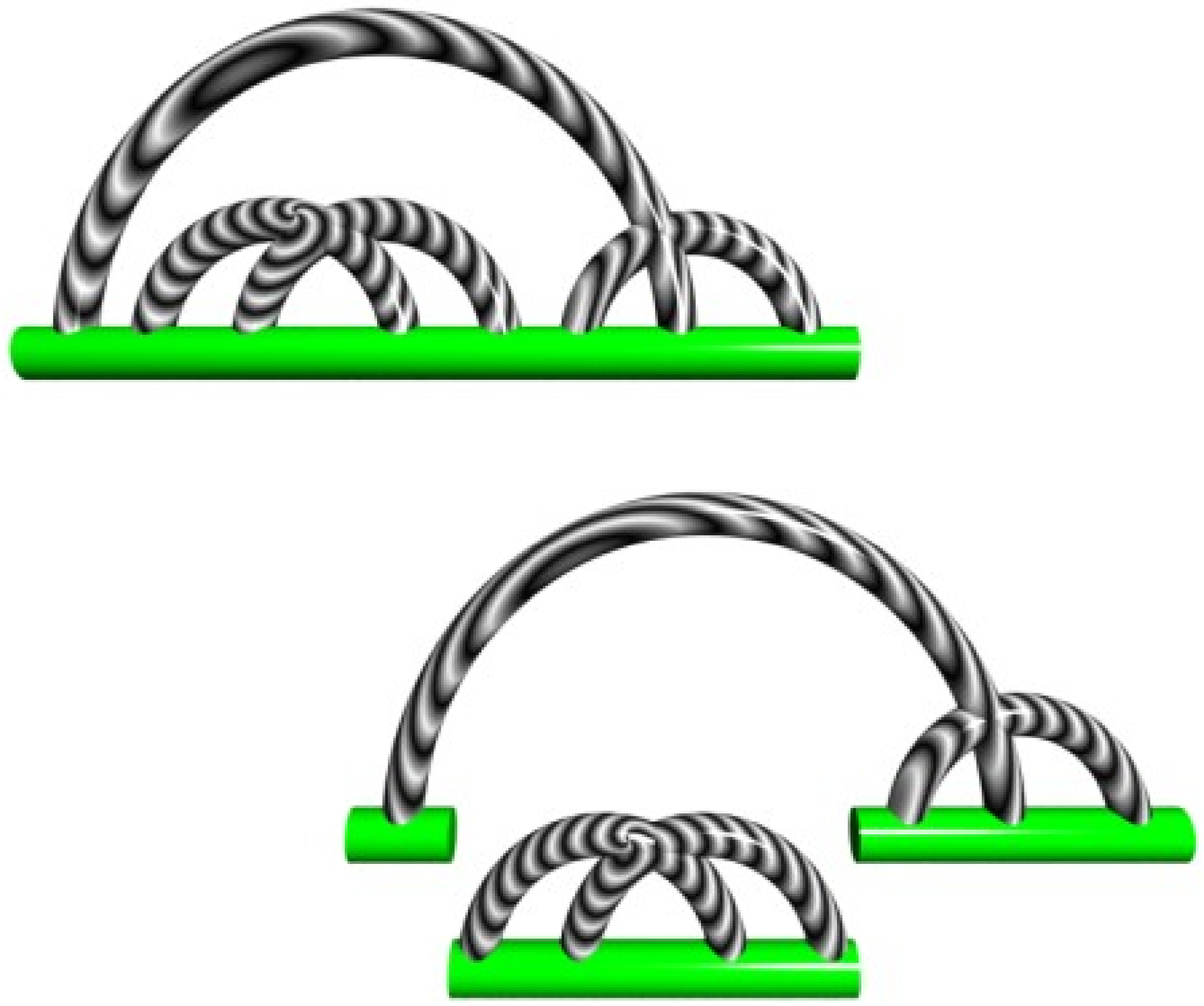}
\put(-135,75){\vector(1,-1){10}} } \framebox{
\includegraphics[width=0.40\textwidth]{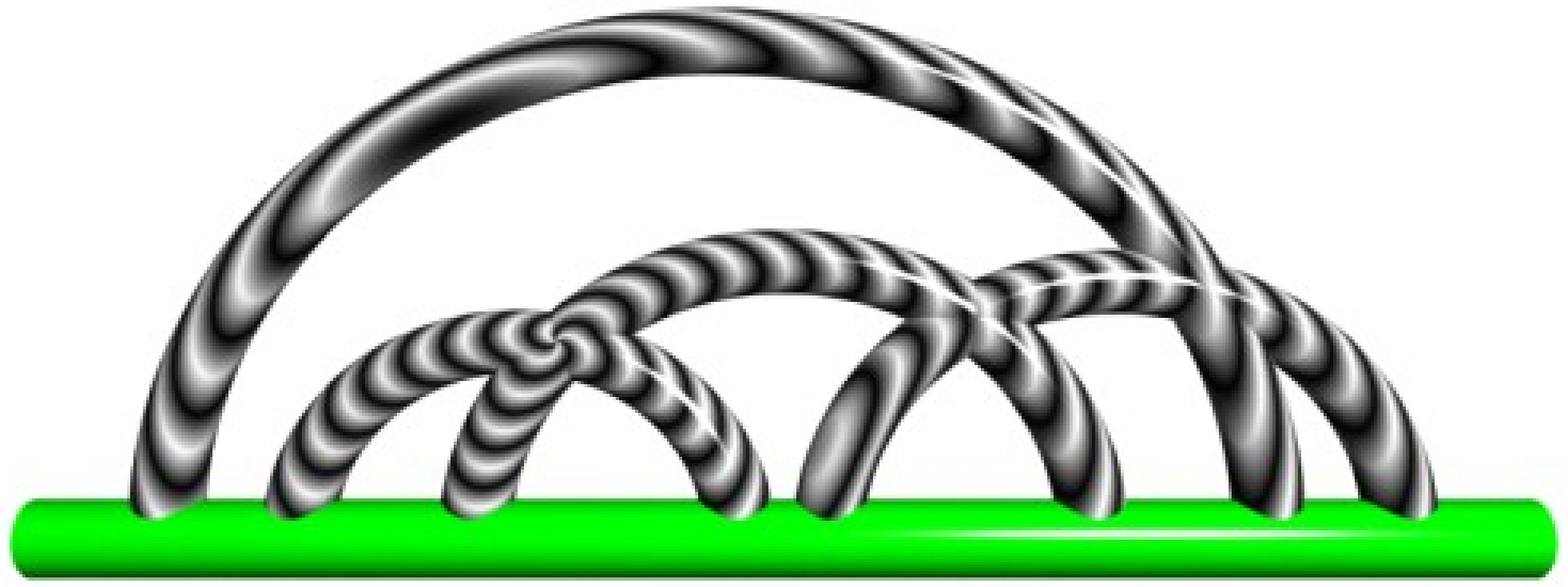}
} \caption{An example of nested diagram (left) and not nested
(right). A nested diagram can be disconnected in two components by
cutting the backbone in two points.} \label{nestedandnot}
\end{figure}

Similarly, if one considers the diagram of fig.\ref{nestedandnot}
(left), its genus is equal to 2. It is composed of an H pseudoknot,
embedded inside another H pseudoknot. A diagram is said to be
embedded or {\it nested} in another, if it can be removed by cutting
two lines while the rest of the diagram stays connected in a single
component. The diagram on the left of fig.\ref{nestedandnot} is
nested, whereas the one on the right is not. It is clear that the
genus of a nested diagram is the sum of the genii of its nested
components. As a result, to any non-nested diagram of genus $g$
there corresponds a nested diagram of same genus, obtained by adding
a pairing line between the first base and the last base of the
diagram. For instance, the 8 diagrams of genus 1 in fig.\ref{genus1}
can be decomposed in 4 non-nested diagrams (left column) and 4
nested diagrams (right column). Therefore, there are only 4
irreducible non-nested diagrams (a,b,c,d) of genus 1. As we shall
see in the next section, pseudoknots (a) and (b) are quite common,
pseudoknot (c) has been seen but is rare, and pseudoknot (d) has not
yet been seen. In the following, a pseudoknot which is irreducible
and non nested is said to be {\it primitive}. Clearly, all RNA
structures can be constructed from primitive pseudoknots. The
primitive diagram for secondary structures is obviously a single
pairing.

\section*{Results and Discussion}
\underline{Analysis of databases}. There are several databases
containing RNA structures. We have analyzed two of them, namely
Pseudobase \cite{Pseudobase} and the wwPDB \cite{PDB} (modulo the
RNAbase database \cite{RNAbase}).

\subsection*{Pseudobase}
\begin{figure}[hbpt]
\centering 
\includegraphics[width=0.48\textwidth]{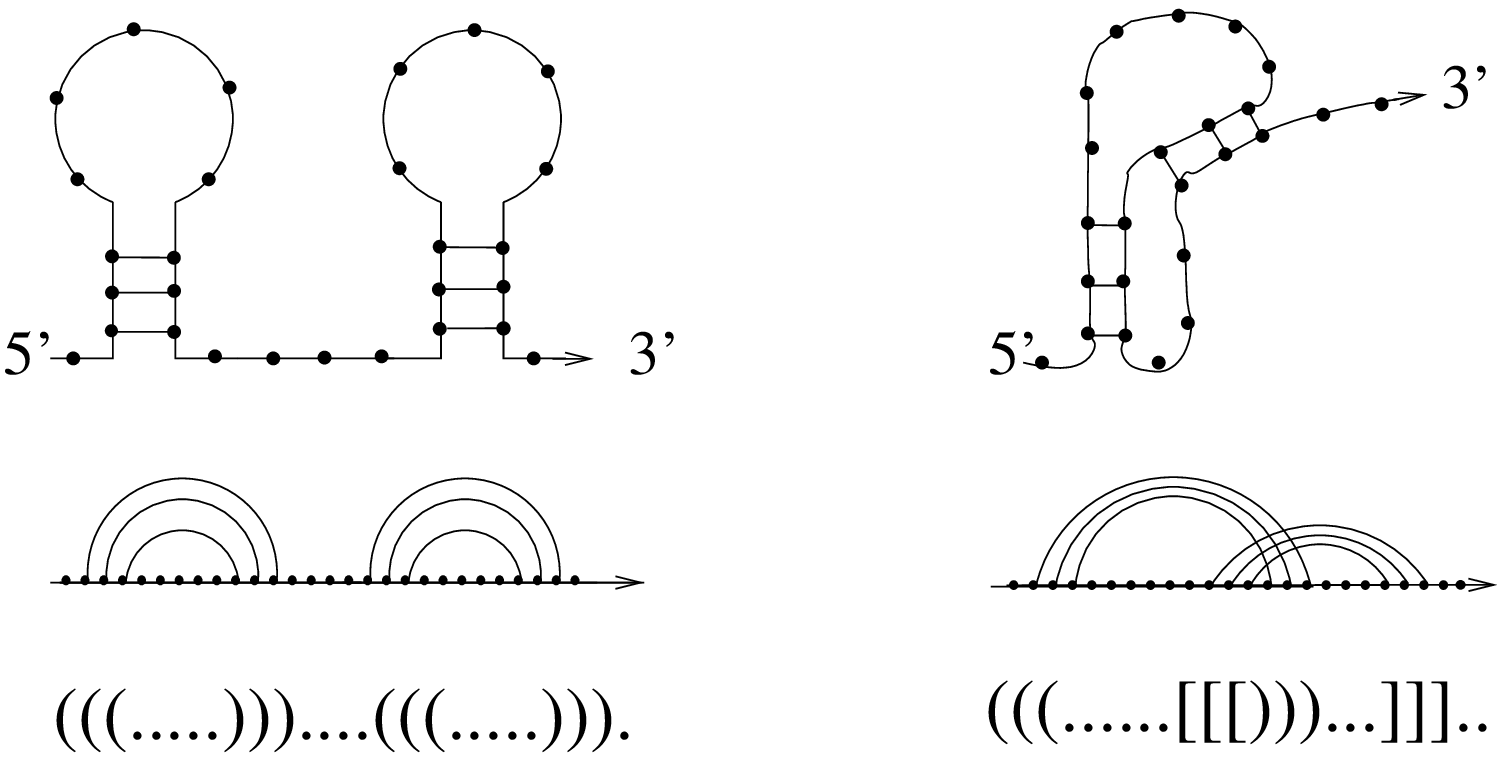}
\caption{The {\it bracket} notation is commonly used for
representing RNA secondary structures with simple pseudoknots. One
stem of the  pseudoknot is represented by parenthesis and brackets
for the other stem. A dot ``." indicates a free base.}
\label{bracket}
\end{figure}

\noindent Pseudobase is a database, containing 246 pseudoknots, at
the time of writing this work. These pseudoknots have been deposited
and validated by several research groups. They are subsegments of
larger RNA sequences, and are displayed in bracket form using
several symbols (see fig.\ref{bracket}). As an example, we show
below one of the pseudoknots from Pseudobase (accession number
PKB210)
\begin{verbatim}
CGCUGCACUGAUCUGUCCUUGGGUCAGGCGGGGGAAGGCAACUUCCCAGGGGGCAACCCCGAACCGCAGCAGCGAC
((((((::(((:::[[[[[[[::))):((((((((((::::)))))):((((::::)))):::)))):))))))::

AUUCACAAGGAA
:::::]]]]]]]
\end{verbatim}
A simple analysis shows that this is an H pseudoknot, of the type
ABAB. Likewise, we analyzed all the 246 pseudoknots of Pseudobase
and found that:
\begin{itemize}
\item there are 238 H pseudoknots (or nested H pseudoknots) of the ABAB
type with genus 1
\item there are 6 kissing hairpin pseudoknots (or nested) of the ABACBC
type with genus 1
\item there is 1 pseudoknot of the type ABCABC (number PKB71) with genus
1
\item there is 1 pseudoknot of the type ABCDCADB (number PKB75) with genus
2
\end{itemize}
Note that the pseudoknot PKB71, from the regulatory region of the
alpha ribosomal protein operon (E.coli organism) is the unique
example of the ABCABC pseudoknot in Pseudobase. Its structure is
\cite{Pseudobase,pkb71}:
\begin{verbatim}
UGUGCGUUUCCAUUUGAGUAUCCUGAAAACGGGCUUUUCAGCAUGGAACGUACAUAUUAAAUAGUAGGAGUGC
(((((((:(((((::::::::[[[[::::[[[[::::{{{{:)))))))))))):::::::::::::::::::

AUAGUGGCCCGUAUAGCAGGCAUUAACAUUCCUGA
:::::::]]]]:::::]]]]:::::::::::}}}}
\end{verbatim}
Its irreducible structure is given in figure \ref{genus1} (third
from the top, on the left). However, looking at sequence alignment,
it is very likely that in fact at least more than 20 other RNA
sequences in the EMBL database \cite{embl} contain pseudoknots of
this kind (A. Mondrag\'on, A. Torres-Larios and K.K. Swinger,
Department of Biochemistry, Molecular Biology and Cell Biology,
Northwestern University, Evanston, IL: {\it private communication}).

\subsection*{The wwPDB databank}
The world wide Protein Data Bank (wwPDB) is a collection of
databases comprising mostly crystallographic and NMR structures of
proteins \cite{PDB}. In addition, as of today, it contains
approximately 850 structures containing at least one RNA molecule.
Among these structures, there are about 300 single RNA structures,
200 containing several RNA fragments, 30 RNA/DNA complexes, 250
RNA/protein complexes and 60 transfer RNA.

Among these 850 structures, there are about 650 structures which have obviously genus 0 (very short sequences, or single or double stranded RNA helices). The number of bases ranges from 22
(2g1w.pdb) which is an H pseudoknot, to 2999 (chain 3 of 1s1i.pdb) which has genus 15.

We have analyzed the remaining 200 structures according to the following scheme:
\begin{itemize}
\item removal of non RNA molecules and extraction of the molecule of interest
\item search for all pairings using the program RNAview
\item selection of relevant pairings (Watson-Crick and G-U wobble)
\item computation of the genus of the corresponding diagram
\end{itemize}

Our  results can be summarized in the following way
\begin{itemize}
\item Transfer RNAs, which are among the smallest RNAs (length of 78),
are made of a single primitive pseudoknot (irreducible and
non-nested) of genus 1 (a kissing-hairpin) nested inside an arch
(see fig.\ref{trna})
\begin{figure}[hbpt]
\centering \framebox{
\includegraphics[width=0.30\textwidth]{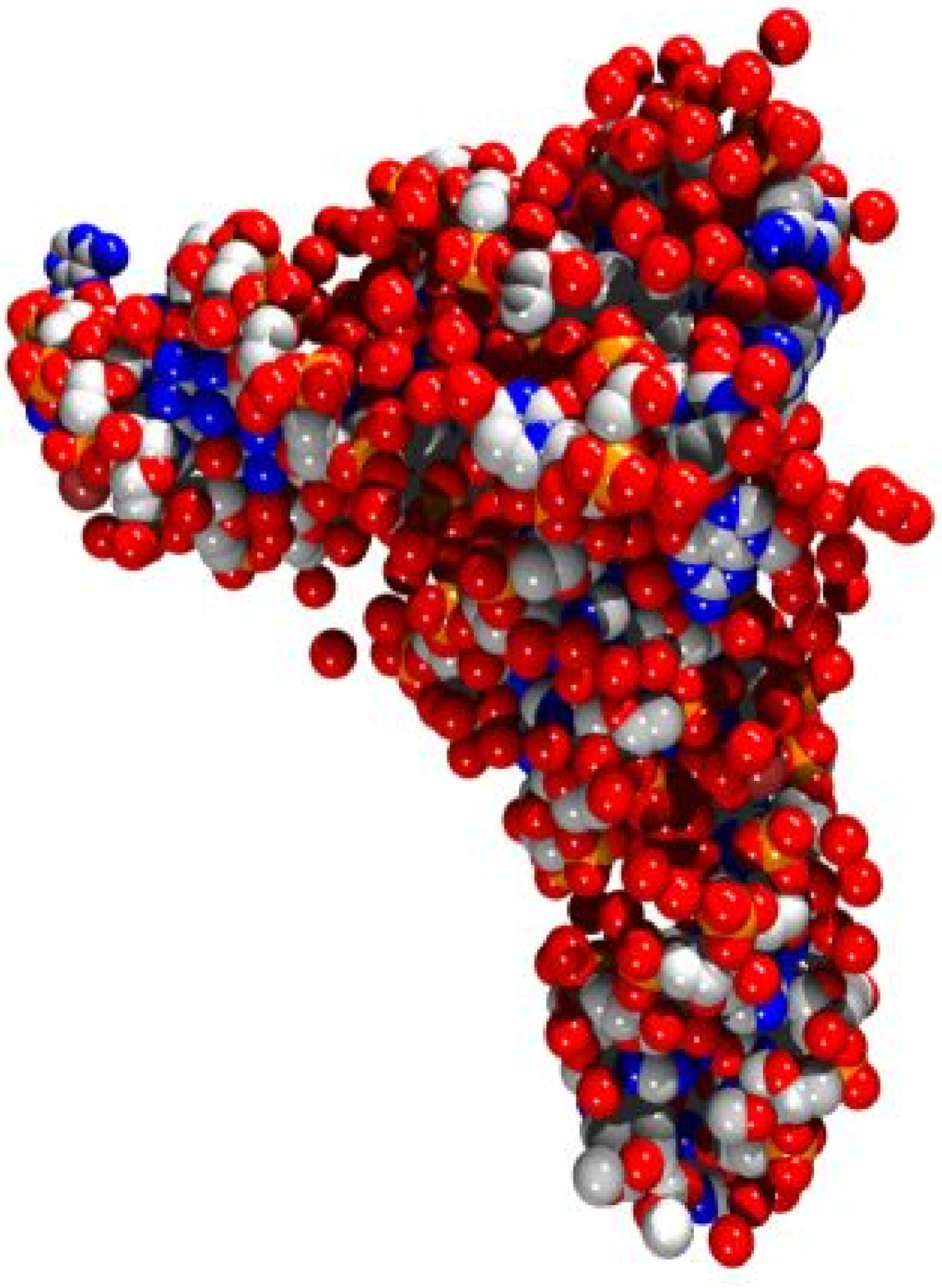}
\includegraphics[width=0.40\textwidth]{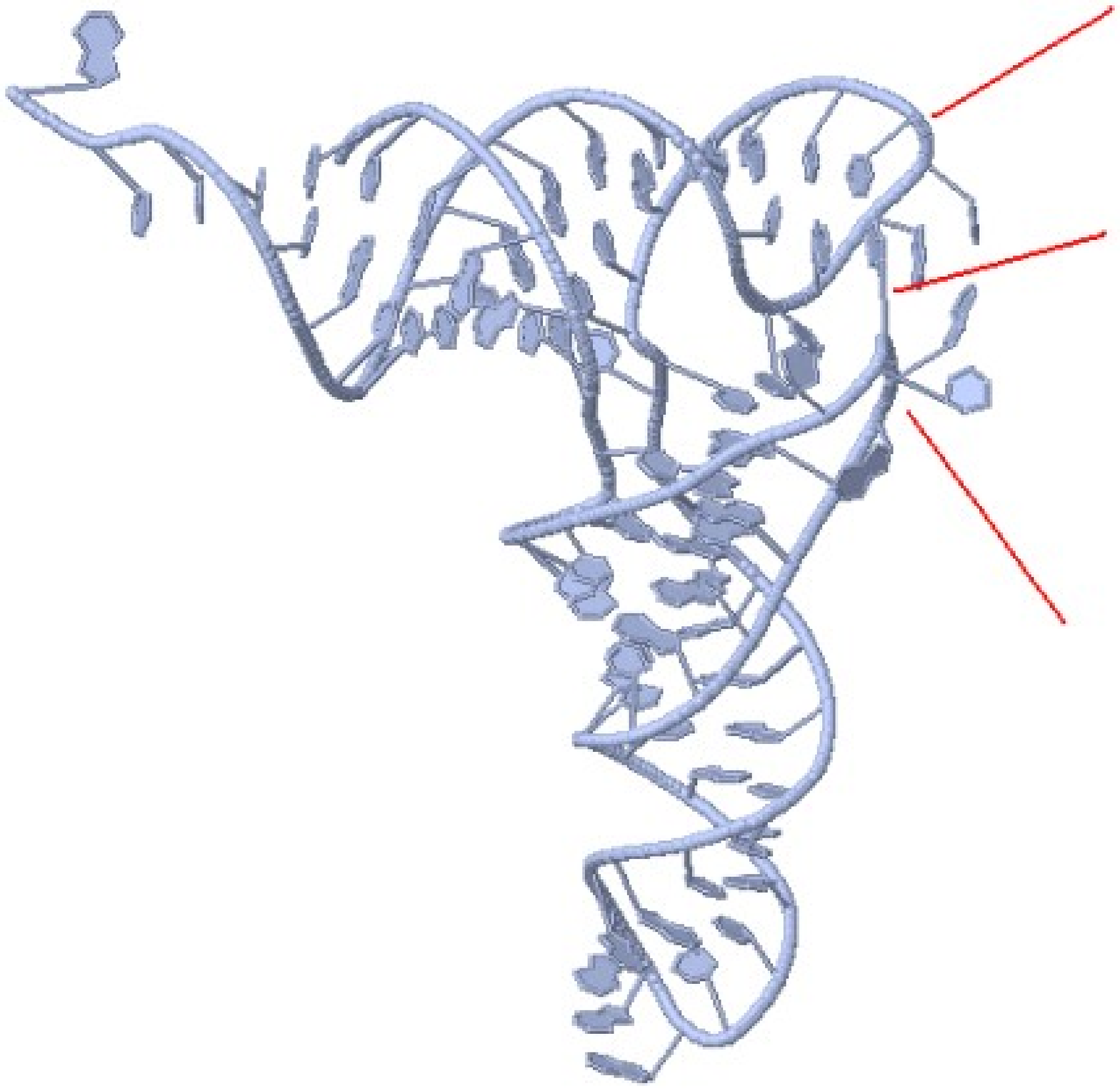}
\put(-23,80){\tiny Hairpin} \put(-20,175){\tiny Hairpin}
\put(-20,140){\tiny Kissing} } \caption{A typical tRNA (PDB 1evv,
\cite{pdb_1evv}). It has the genus 1 of a kissing hairpin
pseudoknot.} \label{trna}
\end{figure}

\item Larger RNAs, such as RNA ribosomal 50s subunits (length larger than 2000),
have total genii less than 18. For an RNA with a non designed random
sequence of length $L$ and without steric constraints, the typical
genus should be $L/4$ \cite{voz_prl}, which in the present case
would be around 500. Even by including steric constraints
\cite{vroz}, the genus would be around $2000\times0.14 \simeq 280$.
In addition, if we analyze these sequences in terms of primitive
pseudoknots, we find that most of the structures are built from very
simple primitive blocks, with genii 1 or 2, nested inside a more
complex pseudoknot, of genus smaller than 8. In fig.\ref{genre1}, we
show an RNA of genus 7 and of length 2825 (the B chain of 1vou.pdb
\cite{1vou_1vp0}) made of 3 H-pseudoknots, 3 kissing hairpins,
nested inside a large kissing hairpin. In fig.\ref{genre4}, we
display an RNA of genus 9 and of length 2825 (the B chain of
1vp0.pdb, of the 50s subunit of E.Coli \cite{1vou_1vp0}), which is
made of 3 H-pseudoknots and 2 kissing hairpins, nested in a
primitive pseudoknot of genus 4.

\begin{figure}[hbpt]
\centering \framebox{
\includegraphics[width=0.44\textwidth]{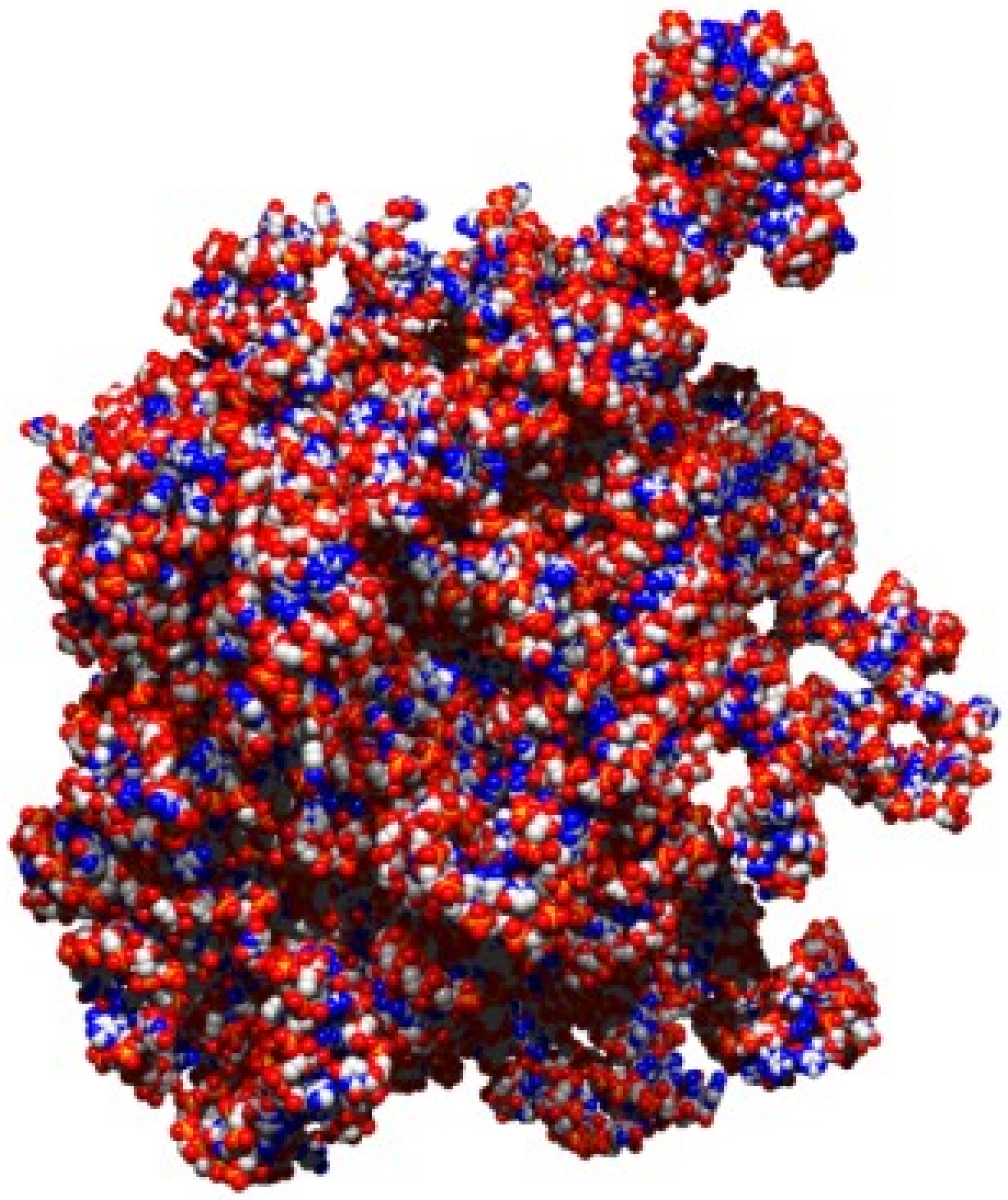}
} \framebox{
\includegraphics[width=0.44\textwidth]{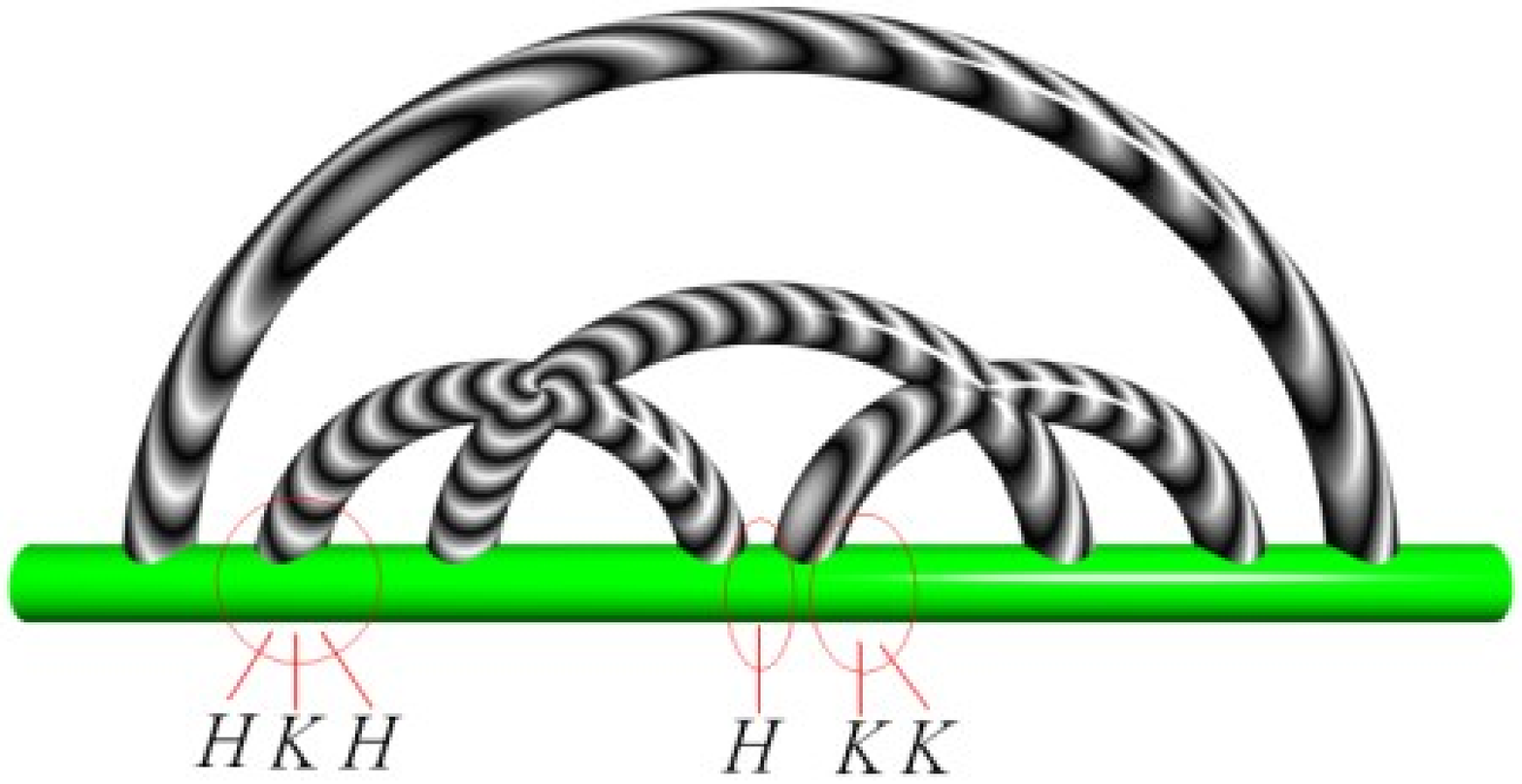}
} \caption{The B chain of PDB 1vou is an RNA of genus 7 and of
length 2825 bases. On the right, the outermost primitive arc
structure is the pseudoknot type b) of the second column in
fig.\ref{genus1}, which has genus 1. Such a primitive structure is
decorated by 6 additional simple pseudoknots of type $H$ and $K$
(type a) and b) in the first column of fig.\ref{genus1},
respectively).} \label{genre1}
\end{figure}

\begin{figure}[hbpt]
\centering \framebox{
\includegraphics[width=0.45\textwidth]{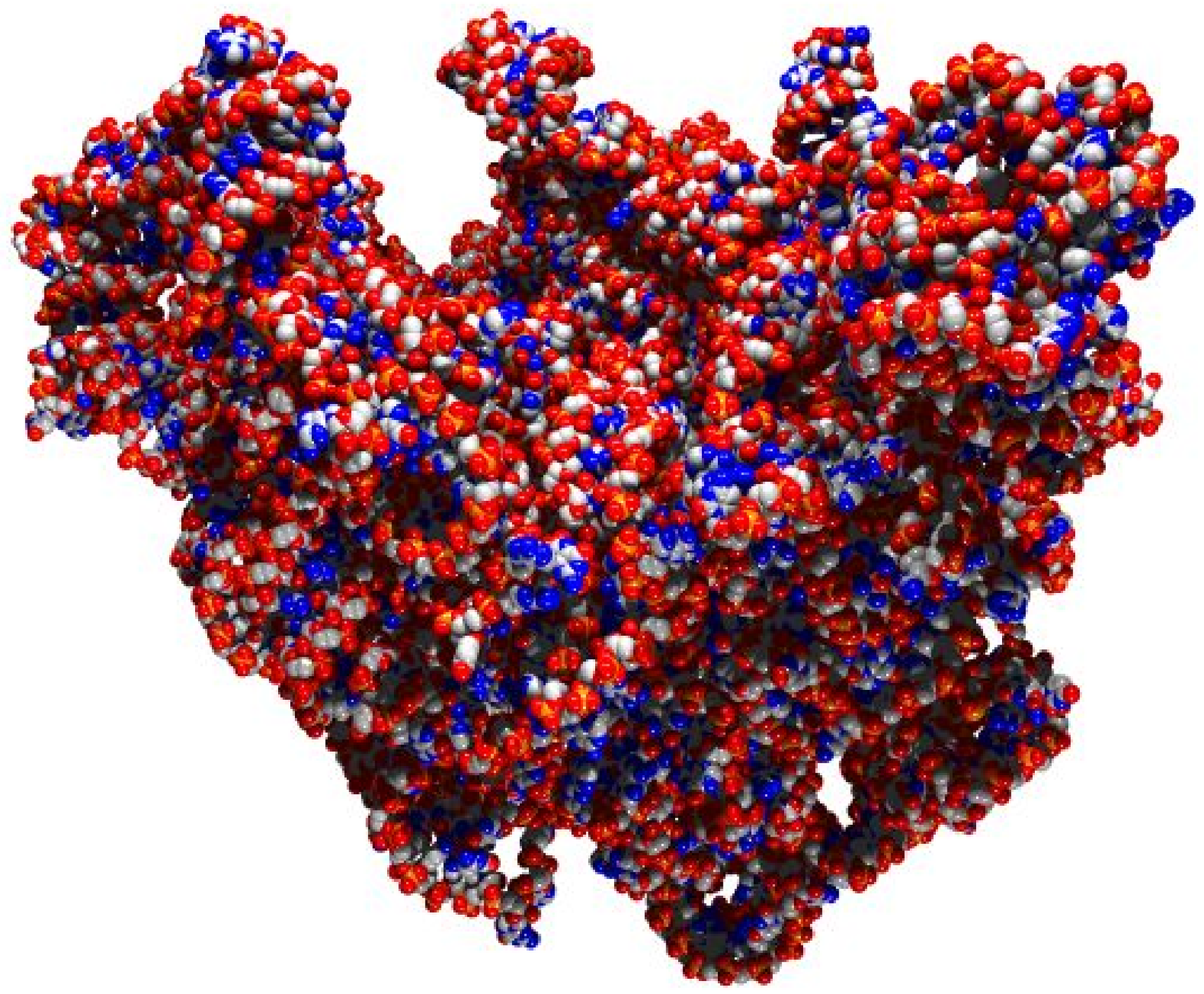}
} \framebox{
\includegraphics[width=0.45\textwidth]{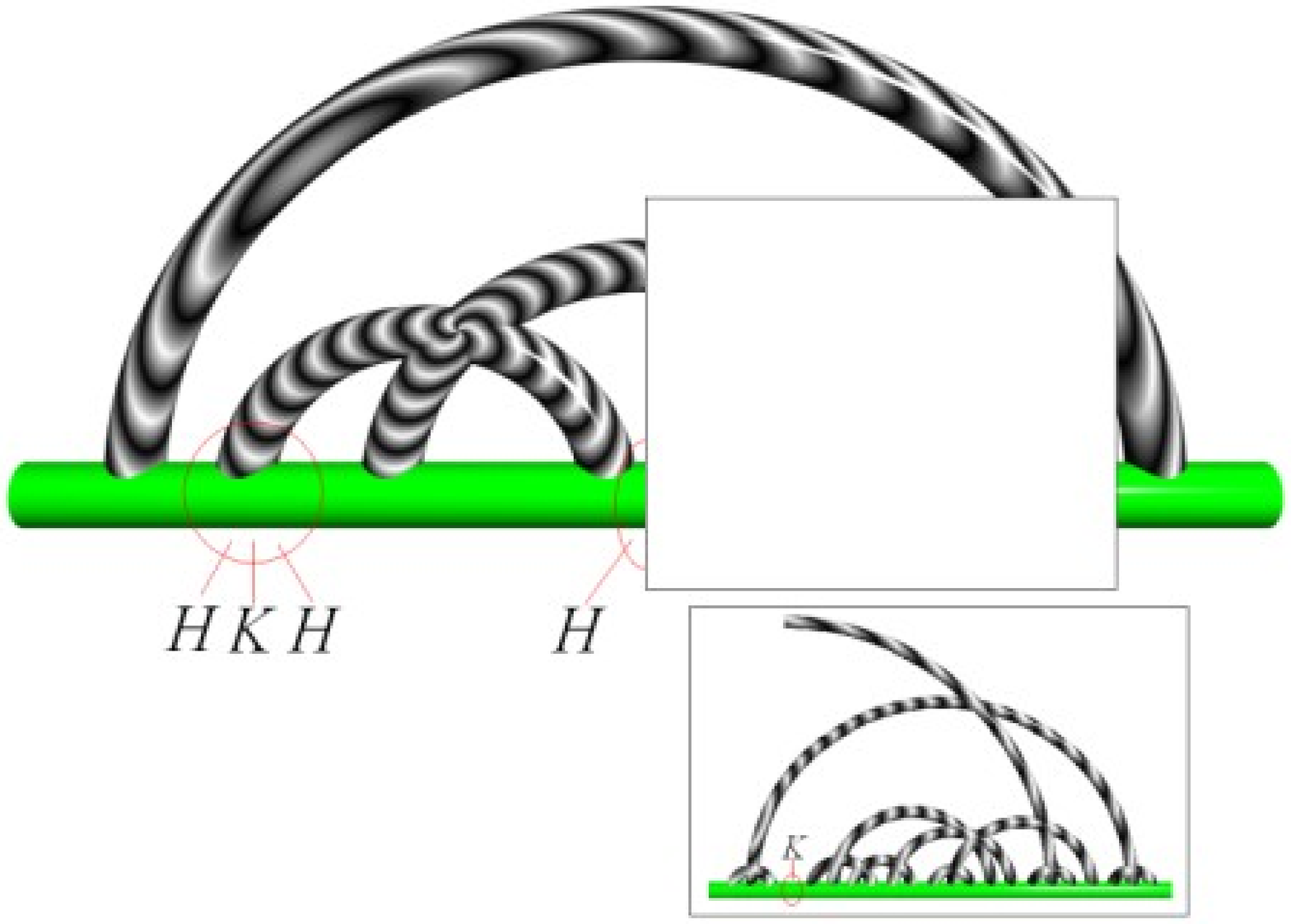}
\put(-95,70){\vector(0,-1){10}} } \caption{The B chain of PDB 1vp0
is an RNA of genus 9 and of length 2825 bases. The outermost
primitive structure is similar to the one of fig.\ref{genre1}, with
a more complex decoration on the right-hand part. There, a complex
pseudoknot with genus 4 is included. Five simple $H$ and $K$
pseudoknots complete the full decoration.} \label{genre4}
\end{figure}

\item There is no hierarchical nesting of the pseudoknots:
The general structure observed in all RNAs of the PDB is that of
several low genus primitive pseudoknots in serie, nested inside a
possibly higher genus ``scaffold pseudoknot". We show in
fig.\ref{genre1} one example of decomposition of a structure
(1vou.pdb, which is a 30s subunit of E. Coli).

\item In fig.\ref{distr1} (left),
we plot the distribution of genii as a function of
the length of the RNA. As mentioned before, the genii are much
lower than what is expected for random sequences, and this is a
manifestation of the specific design of RNA.

\item In fig.\ref{distr1} (right), we plot a
histogram of the statistics of primitive pseudoknots in the PDB. We
see that the genus of primitive pseudoknots is small, typically one
or 2, and that the probability to observe large genii is very small.
This reflects the fact that complex pseudoknots are built from many
small primitive pseudoknots with low genii.

\begin{figure}[hbpt]
\centering \framebox{
\includegraphics[width=0.48\textwidth]{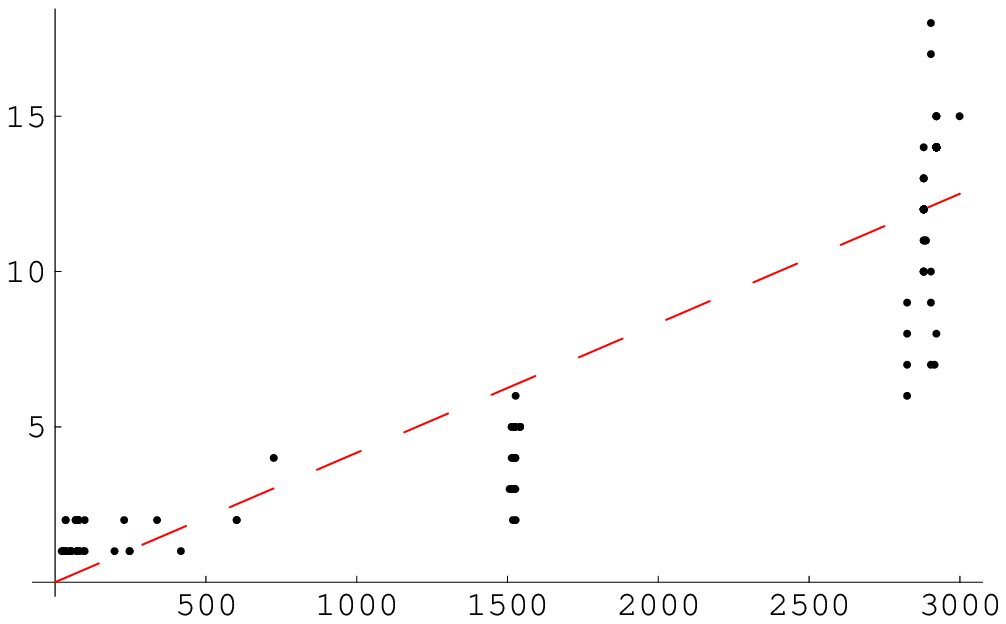}
\put(-228,130){$g$} \put(0,-7){$L$} \quad
\includegraphics[width=0.48\textwidth]{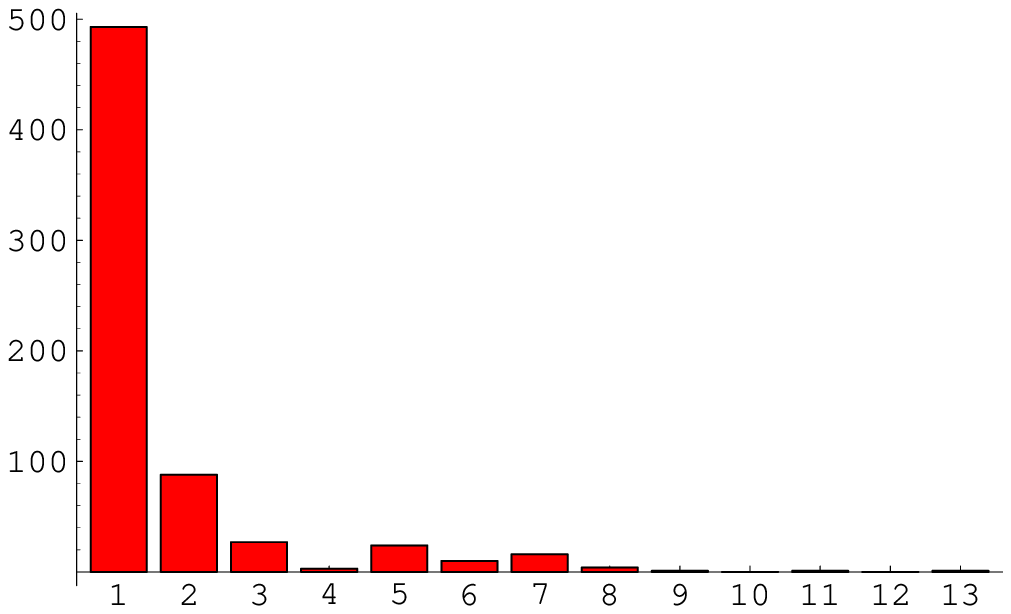}
\put(-239,132){$f$} \put(-5,-5){$L$}} \caption{On the left: total
genus as a function of the number of bases in the RNA molecule. The
interpolating dashed line emphasizes an overall linear behavior. On
the right: histogram distribution of the number $n$ of primitive
pseudoknots as a function of their genus $g$ for all RNA molecules
in the wwPDB database.} \label{distr1}
\end{figure}
\end{itemize}

We conclude by reporting in table \ref{finaltable} the sorted list
of all the PDB files with non-zero genus, according to our
classification. Note that our statistical analysis is biased by the
inherent bias of the PDB: the PDB sometimes contains many structures
of the same molecules, and thus those utilized for the statistical
analysis are not independent.

\section*{Conclusion}
We have shown that RNA structures can be characterized by a
topological number, namely their genus. This genus is 0 for
secondary structures (planar structures), and non zero for
pseudoknots. We have shown how the complexity of the RNA structure
can be analyzed in terms of so-called ``primitive pseudoknots". Any
complex RNA structure can be uniquely decomposed as a sequence of
primitive pseudoknots concatenated sequentially and nested.  A
survey of the existing RNA structures shows that even for large RNA
($\approx$ 3 kb), the genus remains small (smaller than 18), and
natural RNA have a genus which is much smaller than that of paired
structures obtained from random sequences. By capturing the
intrinsic complexity of the structure, the genus provides a natural
and powerful classification of RNA. Finally, a statistical study
shows that complex RNA structures are built from low genii primitive
pseudoknots (genii 0, 1 or 2), and that the most complex primitive
pseudoknots have genus 13. In a forthcoming work, we will show how
this concept of genus can be utilized to actually predict the folded
structure of RNA molecules.

\begin{table}
\scriptsize \centering
\begin{tabular}{|c|l|}
\hline total genus & PDB file accession number \\ \hline
 \hline
 1 & 1b23, 1c0a, 1e8O, 1ehz, 1eiy, 1euq, 1euy, 1f7u, 1f7v,
1fcw,
1ffy, 1fir, 1g59, 1gix-B, 1gix-C, 1grz, \\
  & 1gtr, 1i9v, 1il2, 1j1u, 1jgo-D, 1jgp-D, 1jgq-D, 1kpd, 1kpy, 1kpz, 1l2x, 1l3d, 1mj1, 1mzp,
  1n77, \\
  & 1o0b, 1o0c, 1qf6, 1qrs, 1qrt, 1qru, 1qtq, 1qu2, 1qu3, 1ser, 1sz1, 1tn2,
1tra 1ttt 1u6b-B, 1x8w,\\
 & 1yfg, 1yg3, 1ymo, 1zzn-B, 2a43, 2a64,
2csx, 2fk6, 2g1w, 2tpk, 2tra, 437d, 4tna,
4tra, 6tna,\\
 &  1asy-R, 1asy-S, 1asz-R, 1asz-S  \\
\hline

2 & 1cx0, 1ddy, 1drz, 1et4, 1exd, 1ffz, 1fg0, 1fka, 1pnx, 1sj3,
1sj4, 1sjf, 1u8d, 1vbx, 1vby, 1vbz, \\
 &  1vc0, 1vc5, 1vc6, 1vc7, 1y0q, 1y26, 1y27, 1yoq, 2a2e  \\
\hline

3 &  1i97, 1n34, 1s1h, 1voz, 1yl4-A \\
\hline

4 & 1ibm, 1fjg, 1hnw, 1hnx, 1hnz, 1hr0, 1i95, 1ibk, 1ibl, 1n32,
1n33, 1q86 1vov, 1vox, 1xmo, \\
 & 1xmq-A, 1xnr, 1j5e\\
\hline

5 & 1i94, 1i96, 1n36, 1voq, 1vos, 1xnq, 2avy, 2aw7, 2aw7-A \\
\hline

6 & 1pns, 1voy-B\\
\hline

7 & 1c2w, 1vou-B, 1yl3-A \\ \hline

8 & 1ffk-0, 1vow-B \\
\hline

9 & 1vp0-B, 2aw4-B\\
\hline

10 & 1njm, 1njn, 1njo, 1njp, 2awb-B \\
\hline

11 &    1k01, 1p9x, 1pnu, 1pny \\
\hline

12 & 1j5a, 1jzx, 1jzy, 1jzz, 1nwx, 1nwy-0, 1sm1-0, 1xbp-0, 1y69-0 \\
\hline

13 & 1nkw-0, 1ond, 2d3o \\
\hline

14 & 1jj2, 1k73, 1k8a-A, 1k9m-A, 1kc8, 1kd1, 1kqs-0, 1m1k, 1m90,
1n8r, 1nji, 1q7y, 1q82, 1qvf-0,\\
&  1s72, 1vq4-0, 1vq5-0, 1vq7-0, 1vq8-0, 1vq9-0, 1vqk, 1vql, 1vql-0,
1vqm, 1vqn, 1vqo-0, 1vqp-0, \\
 & 1yhq-0, 1yi2-0, 1yij-0, 1yit-0, 1yj9-0, 1yjn-0, 1yjw-0, 2aar \\
\hline

15 & 1q81, 1qvg, 1s1i-3, 1vq6-0 \\
\hline

16 & -  \\
\hline

17 &   2aw4-B \\
\hline

18 & 2awb-B \\
\hline
\end{tabular}
\caption{List of the PDB files we considered in this paper,
according to their total genus. The notation $xxxx-y$ indicates the
chain number $y$ in the PDB file accession number $xxxx$.}
\label{finaltable}
\end{table}

\section*{Acknowledgements}
\noindent This work was supported in part by the National Science
Foundation under Grant No. PHY 99-07949 and  Grant No. DMR 04-14446,
and by the European program MEIF-CT-2003-501547. G.V. acknowledges
Professor Monica Olvera de la Cruz (Northwestern University) for
support and stimulating discussions.


\end{document}